\documentclass{article}

\usepackage[OT1]{fontenc} 
\usepackage{graphicx}
\usepackage{amssymb,amsfonts,amsmath}
\usepackage{color}

\graphicspath{{Images_original/Preparation_Fig_article/},{Fig_article/}}

\begin{document}

\title{The mechanism of porosity formation during solvent-mediated
phase transformations}

\author{Christophe Raufaste$^{1,2}$ 
,
    Bj\o rn Jamtveit$^{1}$
,
    Timm John$^{1,3}$
,\\ 
    Paul Meakin$^{1,4,5}$
,
 Dag Kristian Dysthe$^{1}$
 }


\maketitle

\begin{abstract}
Solvent-mediated solid-solid phase transformations often result in the formation of a porous medium, which may be stable on long time scales or undergo ripening and consolidation. We have studied replacement processes in the KBr-KCl-H$_2$O system using both \emph{in situ} and \emph{ex situ} experiments. The replacement of a KBr crystal by a K(Br,Cl) solid solution in the presence of an aqueous solution is facilitated by the generation of a surprisingly stable, highly anisotropic and connected pore structure that pervades the product phase. This pore structure ensures efficient solute transport from the bulk solution to the reacting KBr and K(Br,Cl) surfaces. The compositional profile of the K(Br,Cl) solid solution exhibits striking discontinuities across disc-like cavities in the product phase.
Similar transformation mechanisms are probably important in controlling phase transformation processes and rates in a variety of natural and man-made systems.
\end{abstract}

\footnotetext[1]	{Physics of Geological Processes (PGP), University of Oslo, P.O. Box 1048, Blindern, N-0316 Oslo, Norway} 
\footnotetext[2] {Laboratoire de Physique de la Mati\`{e}re Condens\'{e}e, Universit\'{e} de Nice-Sophia Antipolis, CNRS, UMR 6622 Parc Valrose, 06108 Nice cedex 2, France} 
\footnotetext[3] {Instit\"{u}t f\"{u}r Mineralogie, University of M\"{u}nster, Corrensstrasse 24, 48149 M\"{u}nster, Germany.}
\footnotetext[4] {Center for Advanced Modeling and Simulation, Idaho National Laboratory, Idaho Falls, Idaho 83415, USA}
\footnotetext[5] {Institute for Energy Technology, Kjeller, Norway}

\section{Introduction}
\label{Introduction}

The re-equilibration of solids in the presence of fluids is of great importance to the evolution of both natural and synthetic materials. In most cases, the fluid is a liquid solvent, but vapor phase transport in air or other gases may also mediate solid-solid phase transformation (Jiang {\it et al.} 2010). Since Cardew \& Davey (1985) presented the first analysis of the kinetics of ``solvent-mediated phase transformations'', it has become clear that such transformations play a key role in the mineralogical evolution of the earth's crust, as well as in a wide range of industrial processes including control of cement hydration (Tzschichholz {\it et al.} 1996) and production of nanostructured devices (Sun \& Xia 2004; Xia {\it et al.} 2009). Pharmaceuticals and explosives often exhibit polymorphism, and they may transform from one crystal structure to another during storage and/or processing (Rodriguez-Hornedo \& Murphy 1999, Dharmayat {\it et al.} 2008, Heinz {\it et al.} 2009, Achuthan \& Jose 1990, Haleblian \& McCrone 1969).
This is important because the solubility, rate of dissolution, bioavailability, melting point and chemical stability of pharmaceuticals depend on the physical form and crystal structure of the pharmaceutical, and because the processing behavior, including compressibility and granular flow, varies from polymorph to polymorph. Similarly, the performance and stability of explosives varies from one polymorph to another.  In most cases, crystallization from a melt or solution leads to growth of the least stable polymorph (Ostwald 1897, Threlfall 2003), which may transform into another polymorph in the presence of a solvent.
 
Polymorphism in calcium carbonate is well known because of its importance in geological systems and biogeochemistry, and it provides a good example of polymorphism and solvent mediated polymorphic transitions in an aqueous mineral system. The thermodynamics, kinetics and chemical mechanisms of solvent mediated transformations between amorphous CaCO$_3$, ikaite, vaterite, aragonite and calcite have been extensively investigated (Bischoff 1968, Kiyoshi Sawada 1997, Mitsutaka Kitamura 2002, Jun Kawano {\it et al.} 2009, for example). The morphologies of the new mineral phases formed via nucleation and growth have been investigated, but this work has been focused on the early stages of growth (small scale morphology) and the effects of organic molecules, in order to better understand the behavior of CaCO$_3$ in biological systems.  
 
Through a variety of recent studies,  the mechanism of ``coupled dissolution-precipitation" (Wang {\it et al.} 1995; Nahon \& Merino 1997, Putnis 2002) or ``microphase-assisted autocatalysis'' (Anderson {\it et al.} 1998) has been verified as a key process during mineralogical transformations under a wide range of geological conditions (Putnis 2002). 
Precipitation of a solid product, which is more stable than the initial phase, removes dissolved reaction products, and this facilitates the transformation by maintaining the solution in an undersaturated state with respect to the initial solid phase. Although most studies have focused on the micrometer to nanometer scale characteristics of the individual interfaces at which phase transformations take place, there is recent evidence that the porosity generation associated with such processes may also be linked to fluid migration and advective solute transport. Hence, the mechanisms of solvent-mediated phase transformations may have implications for geological alteration and metamorphic processes on very large (up to several kilometers) scales (Pl\"{u}mper \& Putnis 2009).  Regional-scale alteration of rocks is often associated with significant mobilization, transport and concentration of elements that otherwise only occur at low abundances, thus solvent mediated phase transformation is ultimately linked with the formation of economic ore deposits (Oliver {\it et al.} 1994).

Very often, the new solid phase grows epitaxially on the old one (Cesare \& Grob{\'{e}}ty 1995, Putnis \& Putnis 2007),  and the replacement is pseudomorphic (the replacement product crystal has the habit of the original crystal instead of its own usual habit).  Although dissolution at one face of the original crystal and nucleation on another face would be a plausible replacement mechanism, pseudomorphism is more naturally explained by the propagation of a reaction front through the original crystal, with dissolution and precipitation occurring simultaneously at the reaction front. Porosity is essential for the advancement of a mineral replacement reaction front. Without porosity, the reacting phase would quickly become ``protected'' from the fluid phase by a layer of impermeable solid reaction product(s), and the dissolution-reprecipitation process would cease. The protection of the original crystal after a thin protective layer has formed is similar to passivation during corrosion. In fact, in some corrosion processes, the corrosion film is porous, and this facilitates rapid corrosion and formation of a thick corrosion product film (Cox \& Yamaguchi 1994). In anodization, a thick, highly structured porous film, which is formed on a metallic substrate via anodic oxidation  (Crouse {\it et al.}  2000, Gong {\it et al.} 2001), grows via dissolution and oxidation at the base of the porous film (Wu {\it et al.} 2007). Under these circumstances, corrosion/anodization becomes a solid-solid replacement process with the formation of a porous reaction product.  Access of fluids to reactive surfaces area may be maintained by reaction-driven fracturing, and/or pore-formation dominated by growth and dissolution processes. Fracturing depends on volume changing reactions generating sufficient stresses to cause fracturing, and it may occur due to expansion or shrinkage (Jamtveit {\it et al.} 2000; Malthe-S\o renssen {\it et al.} 2006; R\o yne {\it et al.} 2008). Pore formation has been documented during replacement processes in many systems, including alkali feldspars (Worden {\it et al.} 1990; David {\it et al.} 1995), KBr-KCl-H$_{2}$O (Putnis \& Mezger 2004; Putnis {\it et al.} 2005) and fluorapatite (Harlov {\it et al.} 2005). Porosity generation by replacement reactions in solution has also been used to produce nanostructured metals for a variety of applications, including catalysis, optics, electronics and biological sensing (e.g. Sun \& Xia 2004). In many cases, the porous replacement product is a transient morphology. The large interfacial energy associated with the porous structure may drive coarsening and porosity reduction. As a result, most evidence of porosity may be erased. However, some residual porosity such as fluid inclusions usually remains. 

Despite the geological, biological and industrial importance of porosity generation by solvent-mediated phase transformation, the mechanisms generating the often highly complex porous structures are still poorly understood. Here we present an experimental study of KBr replacement by KCl in the presence of an aqueous solution. Our approach allows direct imaging of the reacting interface during porosity formation. We show that the interface develops a highly organized structure due to strong coupling between dissolution, precipitation and local transport in the fluid phase. The porosity network that forms during the replacement process is highly anisotropic and it enables transport between the bulk solution and the evolving reactive surface.

\section{Experimental\label{Experimental}}
\begin{figure}[htb!]
\includegraphics[width=120mm]{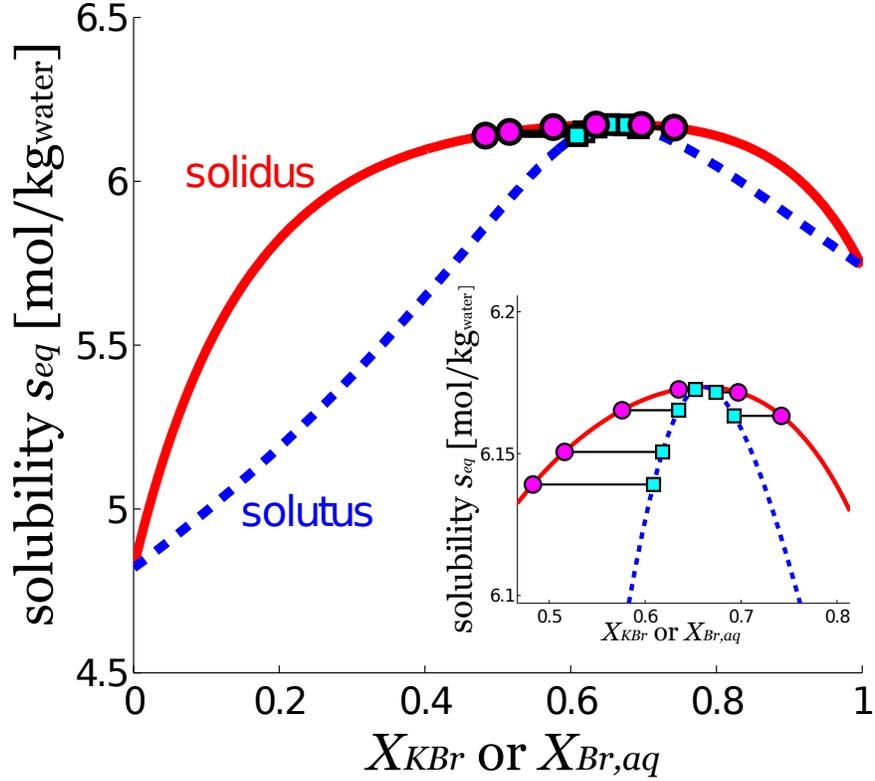}
\caption{Solubility phase diagram of the KBr-KCl-H$_2$O system (modified after Pollok (2004)). The solidus ($s_{eq}$ vs $X_{KBr}$) and solutus ($s_{eq}$ vs $X_{Br,aq}$) curves are shown, where $s_{eq}$ is the solubility of the solution and $X_{Br,aq}$ is the molar proportion of K$^+$Br$^-$ in the solution. The compositions of the different layers ($X_{KBr}^1$, ..., $X_{KBr}^6$) are indicated on the solidus curve. Assuming equilibrium conditions, the local solution compositions, shown in the figure (on the solutus curve), are deduced. Insert: Expansion of the solubility diagram in the region most relevant to the experiments.}
\label{fig:lippmann} 
\end{figure}

We conducted an experimental investigation of the replacement of  KBr crystals by  K(Br,Cl) solid solutions in the KBr-KCl-H$_2$O system. Initially the solution was saturated with respect to KCl only. The composition of the solid solution is described by the formula K[Br$_{X_{KBr}}$Cl$_{1-X_{KBr}}$], where $X_{KBr}$ is the molar proportion of KBr in the solid. The kinetics and thermodynamic properties of this system are well known (Glynn {\it et al.} 1990; Pollok 2004). Fig.~\ref{fig:lippmann} shows the equilibrium relationship between the composition of the solid solution and the composition of the aqueous solution (modified after Pollok (2004)).  The final state of the solid depends on the initial volumes of the solution and of the crystal. Replacement of KBr by K(Br,Cl) leads to a reduction of the total solid volume because 1) at equilibrium, the solubility of K(Br,Cl) is always greater than the initial solution solubility and 2) the molar volume of a K[Br$_{X}$Cl$_{1-X}$] crystal is smaller than that of a KBr crystal (Dejewska 1999) with the same amount of potassium for all $X$ ($0\leq X \leq1$).

The experimental setup allows direct imaging of the interface and its evolution during the reaction.  One side of the crystals was glued to an overlying glass plate with a UV curable adhesive, which prevented reaction from that side and provided a transparent path  to the replacement fronts/interfaces through the glass plate and the KBr crystal (see sketch of the \emph{in situ} optical views of the reaction fronts in Figs.~\ref{fig:optical}{\it BD}). Because the original KBr crystals were optically transparent and isotropic, only the interfaces created optical signals, due to a change in refractive index. The advancing reaction front was continuously imaged \emph{in situ}. The evolution of some crystals was stopped in mid-reaction, and these crystals were cut and polished, and images were obtained using \emph{ex situ} optical microscopy, scanning electron microscopy (SEM),  back-scattered electron (BSE) microscopy and quantitative electron microprobe analysis.

\subsection{In situ experiments} 
KBr crystals (typical dimensions 2x2x2 mm$^3$) were freshly cleaved from initially transparent crystals and immersed in a saturated solution of K$^+$Cl$^-$  with a volume of V=0.54 mL leading to reaction. The saturated KCl solution was obtained by dissolving KCl powder in deionized water. The reaction front was continuously imaged using Olympus BX-60 microscope with 4X-60X  objectives and a Cascade CCD camera (spatial resolution 653x492 pix$^{2}$, intensity resolution 14 bits). White light was used for most imaging and green light ($\lambda_{g}$ = 547 nm) was used for interference pattern quantification. The cavity thickness difference between two fringes is $\lambda_g/(4\cdot n)$, where $n=1.36$ is the refractive index of the solution. Different magnifications were used, ranging from 0.34 to 5.1 $\mu$m/pixel.
\subsection{Ex situ measurements} 
The crystals were removed from the solution, ground roughly under an ethanol solution (a few 100 microns of material was removed to access the inner structure) and immersed in acetone in order to remove all traces of water (these steps were performed within 2 min). Ethanol and acetone do not dissolve the crystals and they dilute the brine solution sufficiently to ensure that no further precipitation or dissolution can occur.
After 12 hours, the crystals were removed from the acetone bath, polished and prepared for analysis. The interface was observed from the side (view 2/xz-plane in Fig.~\ref{fig:optical}) by optical and BSE imaging and quantitative electron microprobe analysis.

\section{Results\label{Results}}

The evolution of the replacement front may be divided into an initial transient stage, which lasts approximately 30 minutes, and a ``steady state'' in which the shapes and the compositions of all structures at the replacement interface remain essentially invariant while the KBr crystal is examined. These recrystallization structures have not been observed before because surface energy driven recrystallization (without chemical change) erases the complex small scale pore structure and leaves the pore geometry observed \emph{ex situ}  by Putnis \& Mezger (2004) when the crystal is removed from the solution bath.  Our approach circumvents this difficulty by  \emph{in situ} observations or by freezing the system before further \emph{ex situ} analysis.
We first describe the striking ``steady state'' geometry and then the initial transient state that casts light on how the recrystallization interface selects the observed steady state.

\subsection{Qualitative optical observations at steady state}
\begin{figure}[htb!]
\setlength{\unitlength}{1cm} 
\centering
\begin{picture}(13.5,13)(0.5,0.0)
\put(0,0){\resizebox{13.5cm}{!}{\includegraphics[width=135mm]{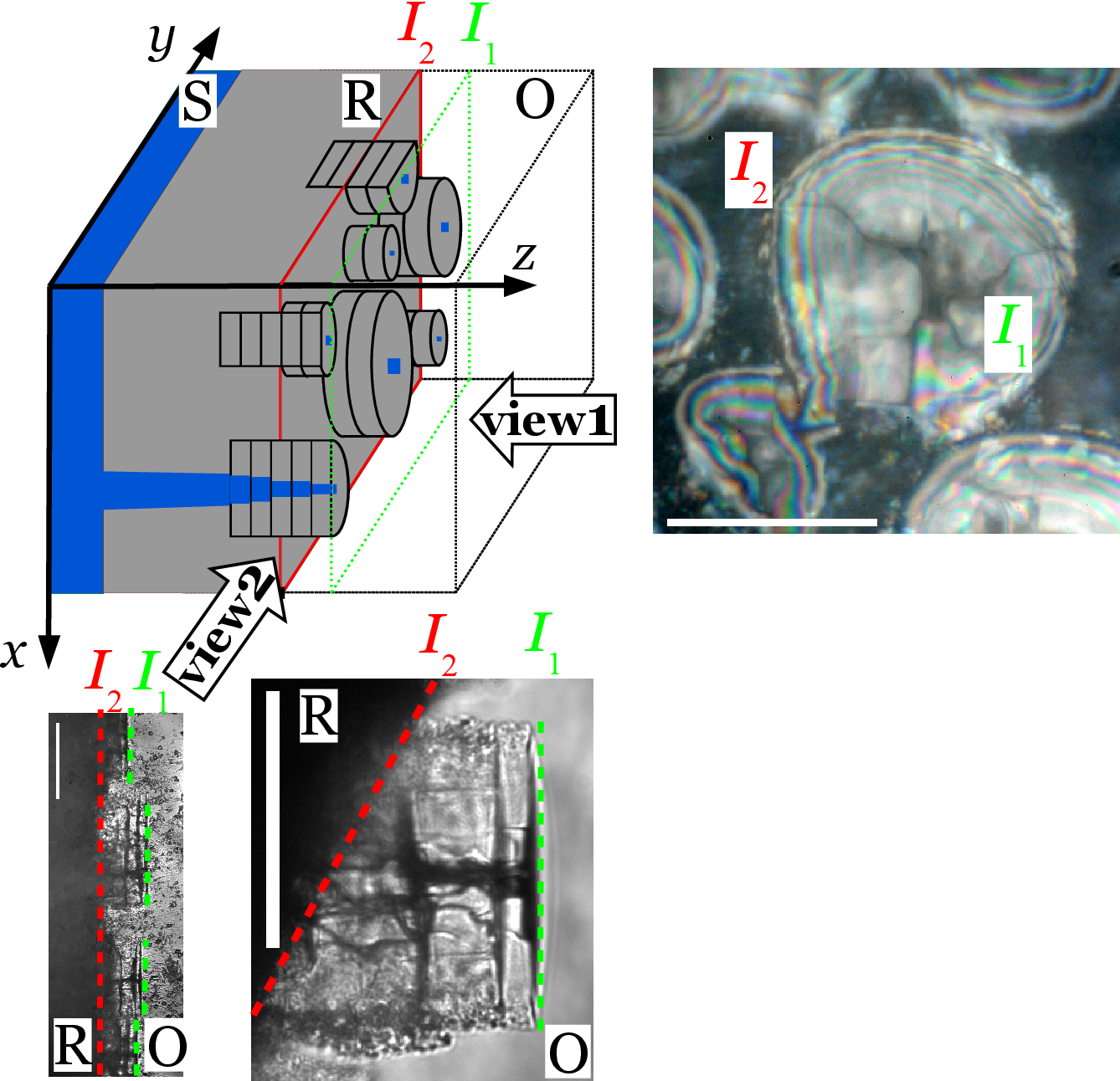}}}
\put(0.1,11.8){\colorbox{white}{A}}
\put(12.9,11.8){\colorbox{white}{B}}
\put(0.1,0.2){\colorbox{white}{C}}
\put(3.1,0.2){\colorbox{white}{D}}
\put(2,7.4){\colorbox{white}{\Large{$\mathbb{C}$}}}
\put(4.3,2.3){\colorbox{white}{\Large{$\mathbb{C}$}}}
\put(11,10.2){\colorbox{white}{\Large{$\mathbb{C}$}}}
\put(4.5,1){\colorbox{white}{\Large{$\mathbb{D}$}}}
\put(5.5,1){\colorbox{white}{\Large{$\mathbb{D}$}}}
\put(6.4,1){\colorbox{white}{\Large{$\mathbb{D}$}}}
\put(7.7,5.3){\colorbox{white}{\large{\bf{S}}: (dark blue) solution}}
\put(7.7,4.6){\colorbox{white}{\large{\bf{O}}: (transparent) original KBr crystal}}
\put(7.7,3.9){\colorbox{white}{\large{\bf{R}}: (light gray) replaced K(Br,Cl) crystal}}
\put(7.7,3.2){\colorbox{white}{\large{\color{green} $I_1$}: advancing boundary}}
\put(8.2,2.7){\colorbox{white}{\large of the leading replacement zone}}
\put(7.7,2){\colorbox{white}{\large{\color{red} $I_2$}: trailing edge}}
\put(8.2,1.5){\colorbox{white}{\large of the leading replacement zone}}
\put(7.7,0.8){\colorbox{white}{\large{$\mathbb{C}$}: pore channel}}
\put(7.7,0.1){\colorbox{white}{\large{$\mathbb{D}$}: disk-like cavity}}
\end{picture}
\caption{Optical images of the replacement interface. 
{\it A}) Representation of the experimental system with solution (S), replaced crystal (R) and original KBr crystal (O). The layered cylinders of organized replacement material extend through the leading replacement zone between the front $I_1$  and the front $I_2$ and into the brine saturated porous K(Br,Cl) trailing replacement zone that contains no pure KBr (the slices on the edges indicate that the cylinders extend into the fully replaced crystal region). If a slice cuts a cylinder along its axis a pore channel ($\mathbb{C}$) connecting $I_1$ to the bulk solution is observed.
The replacement fronts  $I_1$ and $I_2$ are imaged in two orientations: {\bf view 1}: image of the replacement front planes ($I_1$ and $I_2$), looking towards the $x-y$-plane, and {\bf view 2}: image perpendicular to the replacement front planes, looking towards the $x-z$-plane (or $y-z$-plane).  
{\it B}) \emph{In situ} reflected light microscopy color image in view 1 focused on one cylinder top of front $I_1$. 
{\it C}) \emph{Ex situ} optical microscopy image in view 2 showing the original, transparent crystal to the right, the dark, replaced crystal to the left and the two replacement zone boundaries $I_1$ (4 cylinders) and $I_2$.
{\it D}) \emph{In situ} transmitted light microscopy image in view 2 of a similar region. One replacement cylinder is observed, with its pore channel ($\mathbb{C}$) and at least 3 disk-like cavities ($\mathbb{D}$).  In {\it D}) only, the initial crystal was miscut leading to a different orientation for $I_1$ and $I_2$ (see text for explanation). 
Scale bars are 100~$\mu$m.
}
\label{fig:optical} 
\end{figure}

\begin{figure}[htb!]
\setlength{\unitlength}{1cm} 
\centering
\begin{picture}(12,6)(0.0,0.0)
\put(0,0){\includegraphics[width=120mm]{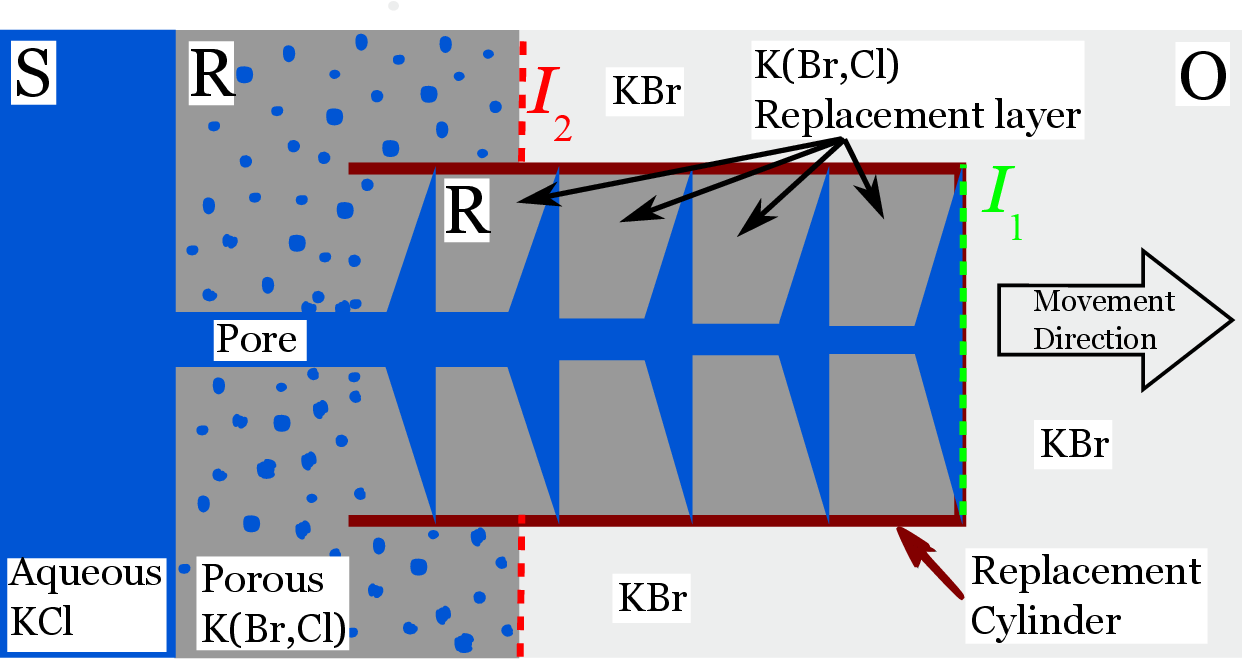}}
\put(2.9,2.95){\colorbox{white}{\large{$\mathbb{C}$}}}
\put(3.6,2){\colorbox{white}{\large{$\mathbb{D}$}}}
\put(4.8,2){\colorbox{white}{\large{$\mathbb{D}$}}}
\put(6.1,2){\colorbox{white}{\large{$\mathbb{D}$}}}
\put(7.4,2){\colorbox{white}{\large{$\mathbb{D}$}}}
\put(8.7,2){\colorbox{white}{\large{$\mathbb{D}$}}}
\end{picture}
\caption{Unscaled diagram in view 2 of the replacement morphology. Colors and symbols refer to the caption of Fig.~\ref{fig:optical}.
}
\label{fig:cylinderrepresentation} 
\end{figure}

Figs.~\ref{fig:optical} and~\ref{fig:cylinderrepresentation} show schematic representations of the experimental system and several images that illustrate its qualitative behavior. 

\begin{figure}[htb!]
\setlength{\unitlength}{1cm} 
\centering
\begin{picture}(13.5,12)(0.5,0.0)
\put(0,0){\resizebox{13.5cm}{!}{\includegraphics[width=135mm]{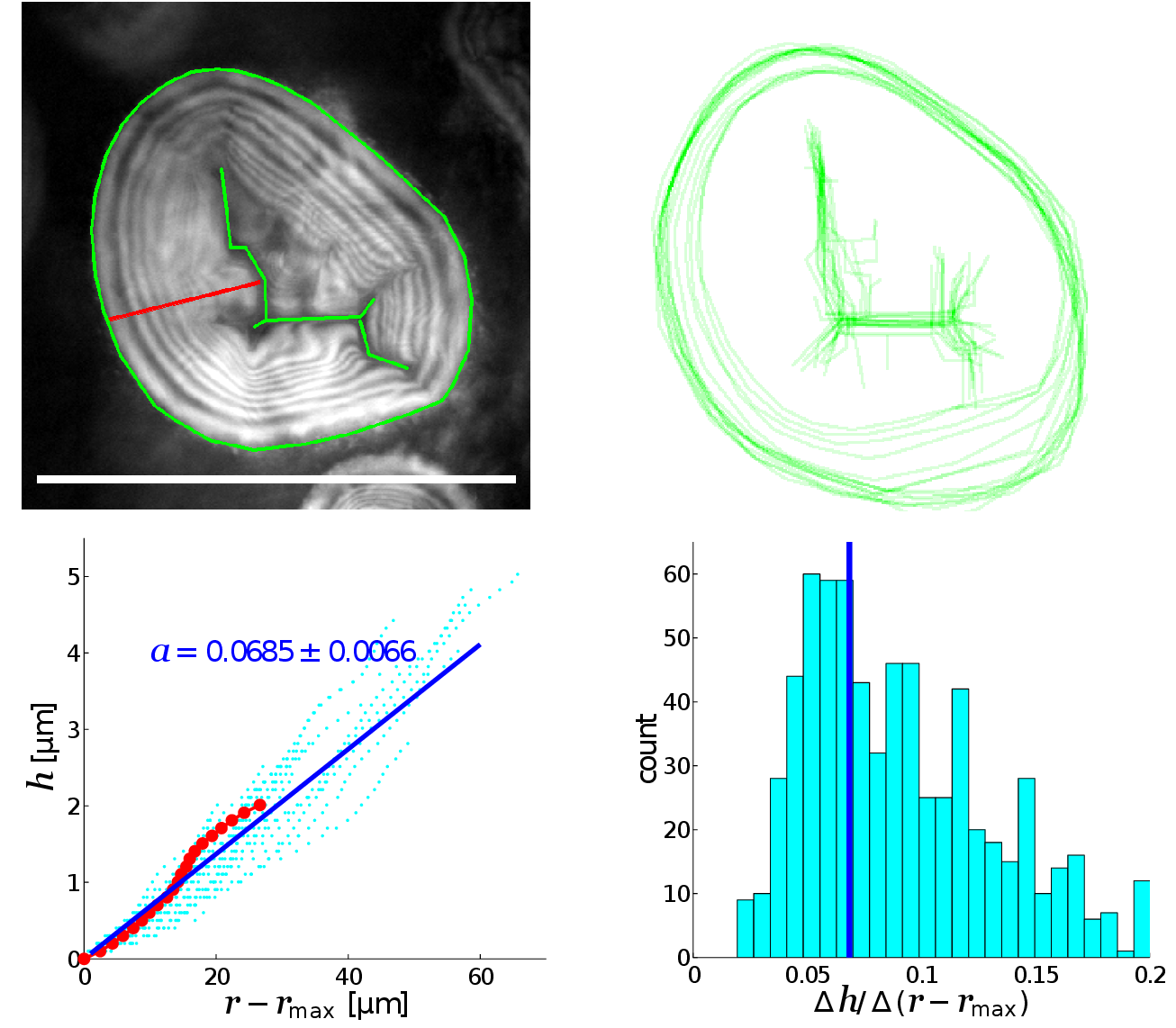}}}
\put(0.2,11.5){\colorbox{white}{A}}
\put(7,11.5){\colorbox{white}{B}}
\put(0.2,5.2){\colorbox{white}{C}}
\put(7,5.2){\colorbox{white}{D}}
\end{picture}
\caption{
Characterization of fluid filled cavity morphology. {\it A}) Optical image, with a green filter, of the first replacement cavity (view 1); the scale bar is 100~$\mu$m. {\it B}) Superposition for different times (between t = 30 and 120 min) of the edges of the cavity and the central channel (green in {\it A})). {\it C}) Reconstruction of the thickness profile from optical images over different cavities and times (the profile corresponding to picture {\it A}) is displayed in red). A linear interpolation leads to an average slope value $a = 0.0685$. This value gives the aspect ratio between the thickness and the radius of the cavity. {\it D}) statistics of the local slopes.  The average slope, $a$, is displayed in dark blue.} 
\label{fig:interface} 
\end{figure}

The pictures show that the interface between the unaltered KBr crystal and the porous K(Br,Cl) saturated with KBr/KCl/H$_2$O solution has a complex geometry. This interface is bounded by two planar replacement fronts, $I_1$ and $I_2$. $I_1$ separates the region that contains only KBr from the region that also contains K(Br,Cl) and solution, and $I_2$ separates the region that contains only K(Br,Cl) and solution from the region that also contains some unaltered KBr (Fig.~\ref{fig:optical}{\it ACD}). 
The region between planes $I_1$ and $I_2$ is the ``leading replacement zone'' in which KBr is replaced with K(Br,Cl).  In this zone, replacement occurs inside distinctive, generally unconnected, more-or-less cylindrical structures that extend from front $I_1$ to front $I_2$, and often some distance beyond $I_2$ (Fig.~\ref{fig:optical}{\it A}). Within the leading replacement zone, the material between the cylinders is unaltered KBr, which has never been in contact with the solution (as Fig.~\ref{fig:lippmann} indicates, once KBr has dissolved in KBr/KCl/H$_2$O, it cannot be reprecipitated as pure KBr).
Observed through the unaltered KBr, from the direction perpendicular to the interfaces (called view 1), the cylinders appear brighter than their surroundings, have cross-sections of varying size ($10^2-10^4 \mu$m$^2$) and occupy approximately 60\% of the plane in which they lie (Fig.~\ref{fig:optical}{\it B}). All the cylinder tops are almost coplanar and correspond to front $I_1$ (the most advanced boundary of the replacement zone). The observation of interference fringes indicates that a thin fluid filled cavity lies at the KBr end of every cylinder (Figs.~\ref{fig:optical}{\it B},~\ref{fig:interface}{\it A}). These fringes (colored in Fig.~\ref{fig:optical}{\it B}, black and white in Fig.~\ref{fig:interface}{\it A}), visible in reflected light due to reflections on both surfaces of the cavities, indicate that the cavities become thinner towards their edges (as Fig.~\ref{fig:optical}{\it D} also clearly shows).

\begin{figure}[h!]
\setlength{\unitlength}{1cm} 
\centering
\begin{picture}(11,17.3)(0.0,0.0)
\put(0,0){\includegraphics[height=173mm]{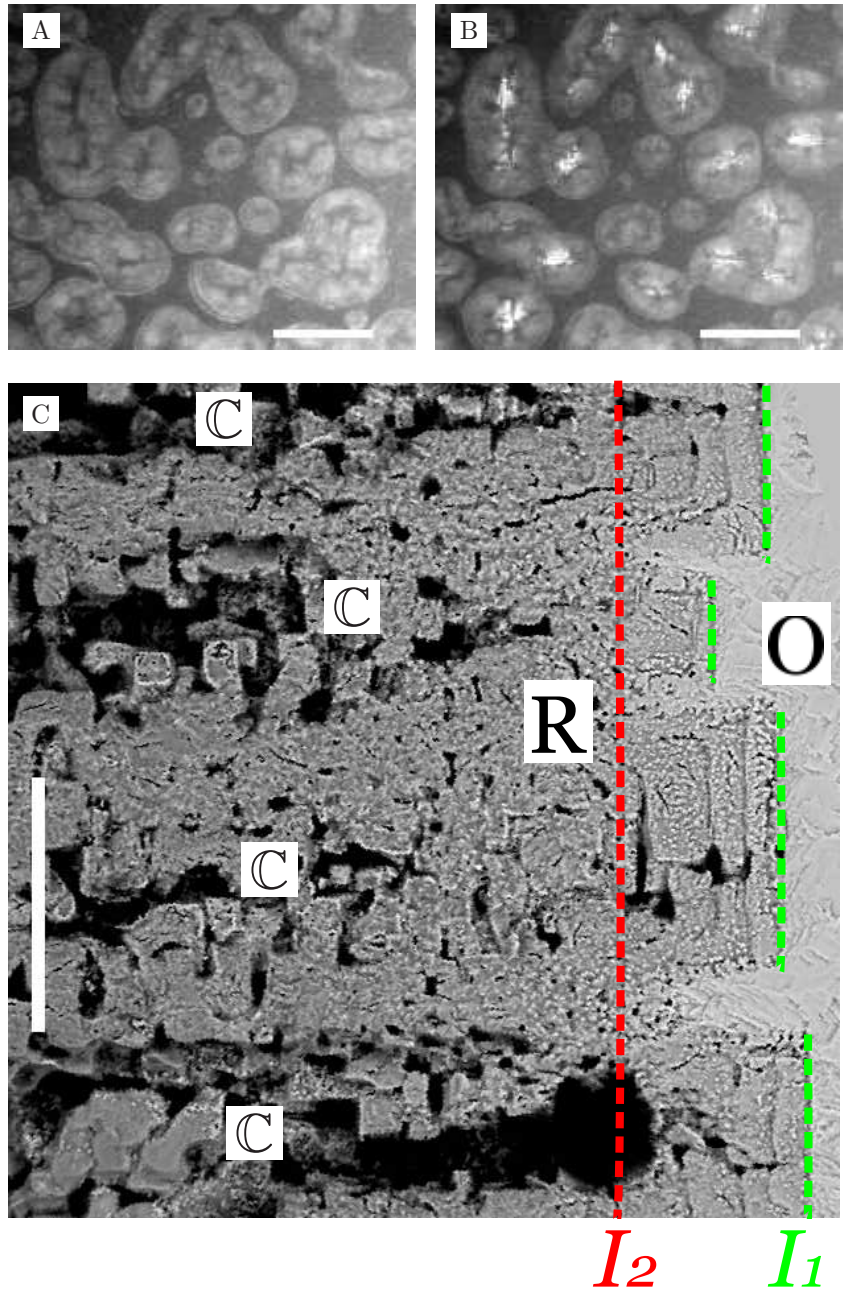}}
\put(0.2,16.8){\colorbox{white}{A}}
\put(5.9,16.8){\colorbox{white}{B}}
\put(0.2,11.7){\colorbox{white}{C}}
\put(2.5,11.5){\colorbox{white}{\huge{$\mathbb{C}$}}}
\put(4.2,9){\colorbox{white}{\huge{$\mathbb{C}$}}}
\put(3.1,5.5){\colorbox{white}{\huge{$\mathbb{C}$}}}
\put(2.9,2){\colorbox{white}{\huge{$\mathbb{C}$}}}
\end{picture}
\caption{Pore channel geometry. {\it A}) \emph{In situ} reflected light microscopy image in view 1 focused on front $I_1$.  
{\it B}) \emph{In situ} reflected and transmitted light microscopy image with the same focal plane and region as {\it A}). Pore channels transmitted light directly through the whole replaced crystal. {\it C}) \emph{Ex situ} BSE microscopy image in view 2. 4 replacement cylinders followed by 4 pore channels (horizontally oriented black shapes) can be seen. Symbols refer to the caption of Fig.~\ref{fig:optical}. Scale bars are 100~$\mu$m.
}
\label{fig:channel} 
\end{figure}

Observed in view 1 the disc-like cavities move towards the imaging objective (into the KBr crystal) while retaining their gross shapes. This translation of the cavities can only occur by dissolution of KBr crystal at their advancing boundaries and precipitation of approximately the same volume of K(Br,Cl) at the opposing interface, which grows the replacement crystal into the advancing disc-like cavity. In the center of each disc there is a dark region (Figs.~\ref{fig:optical}{\it B},~\ref{fig:interface}{\it A},~\ref{fig:channel}{\it A}) with edges that are aligned with the low index (cubic) crystal planes. The cylindrical structures contain newly precipitated crystals organized into several layers separated by brine, which can be seen in views perpendicular to the replacement interface (view 2, Figs.~\ref{fig:optical}{\it CD},~\ref{fig:microprobe}{\it A}). In transmitted light (view 1, Fig.~\ref{fig:channel}{\it B}) the central regions of the cylinders appear brightest, indicating that a fairly straight channel or pore ($\mathbb{C}$) connects the disc-like cavities between the newly precipitated K(Br,Cl) and extends towards the bulk solution. 

Up to 6 layers of K(Br,Cl) per cylinder structure, which extend above and below front $I_2$, were observed (3-4 are visible in the optical images, 6 in the BSE image), and they are separated by disk shaped fluid filled cavities. The first cavity (rightmost vertical dashed lines in Figs.~\ref{fig:optical}{\it CD},.~\ref{fig:microprobe}{\it A}) separates the original crystal from the first layer of the newly formed crystal, and it is the one that is observed in view 1 (Fig.~\ref{fig:optical}{\it B}). All the cavities inside one cylinder are similar in lateral and vertical extent to the first one, and they are organized around the same channel.

The channels ($\mathbb{C}$) can be observed in the \emph{ex situ} image (view 2) when they are cut by the slice plane (Fig.~\ref{fig:channel}{\it C}). They are indicated in the sketchs Figs.~\ref{fig:optical}{\it A},~\ref{fig:cylinderrepresentation}. 

The replacement cylinder is always oriented parallel to one of the high index crystal planes of the original crystal. Thus, when the original KBr crystal is miscut with respect to the high index plane with an angle $\gamma$ the angle between the plane of the individual cavities (or the K(Br,CL) layers) and the front $I_2$ (or the original surface) is also equal to $\gamma$, and the angle between the axis of the central pore and $I_2$ is $\pi/2 - \gamma$.

The pore volume inside each cylinder, created by the solid volume reduction associated with the replacement process, has a strongly anisotropic structure consisting of a central, essentially linear, pore along the axis of each cylinder, which connects an array of disk-like cavities, which are oriented perpendicular to the central pore. The characteristic scales of this surprisingly ordered pore volume are on the order of 10 microns or larger, and the pore can be easily resolved by an optical microscope. The scale of the porosity in the rest of the K(Br,Cl) (outside cylinders) formed by the replacement reaction is typically smaller than 1 $\mu$m, which is too small to be resolved with the present setup. This fine porous material appears in black in Figs.~\ref{fig:optical}{\it BCD},~\ref{fig:channel}{\it AB} , mainly because it strongly scatters light. The replacement material in the ``trailing replacement zone'' behind advancing interface $I_2$ (region R in Fig.~\ref{fig:optical}{\it A}) can only be observed in high resolution \emph{ex situ} images (Fig.~\ref{fig:channel}{\it C}). Dark patches are former fluid filled pores (the fluid is evacuated during processing) that have a much more random morphology than the channels ($\mathbb{C}$), and it can be seen that traces of the succession of replacement layers remain some distance behind interface $I_2$ in region R. 

The distance between the two replacement interfaces $I_1$ and $I_2$ increases slowly as they advance through the crystal, and the interface seems to translate as a whole in time lapse videos (Video S1).

\subsection{Quantitative measurements at steady state}
\begin{figure}[htb!]
\setlength{\unitlength}{1cm} 
\centering
\begin{picture}(13.5,14)(0.5,0.5)
\put(0,0){\resizebox{13.5cm}{!}{\includegraphics[width=135mm]{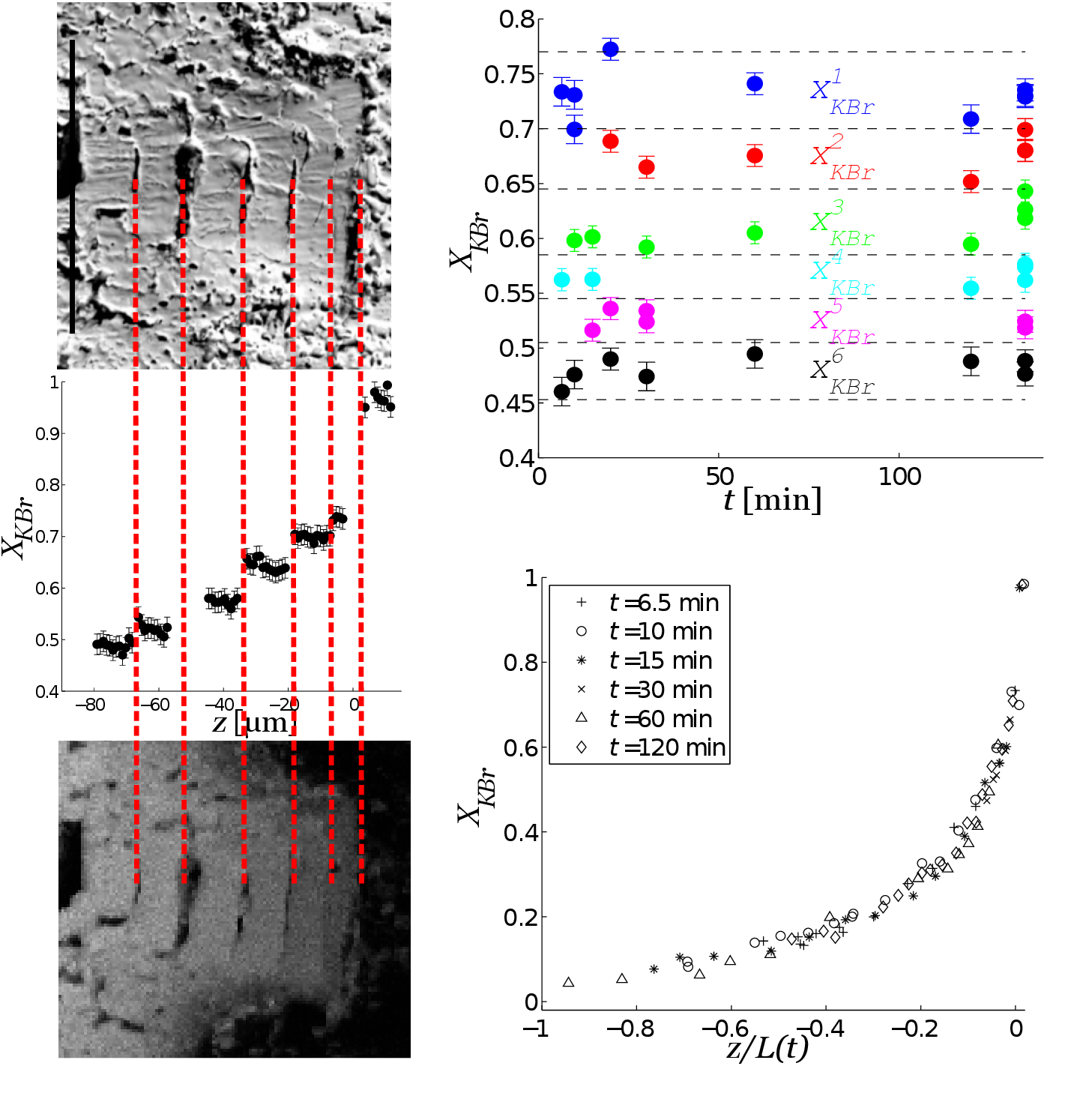}}}
\put(1,13.7){\colorbox{white}{A}}
\put(1,9){\colorbox{white}{B}}
\put(1,4.4){\colorbox{white}{C}}
\put(5.6,13.7){\colorbox{white}{D}}
\put(5.6,6.5){\colorbox{white}{E}}
\put(13.5,14){\colorbox{white}{.}}
\end{picture}
\caption{Solid solution composition measurements. {\it A}) BSE image of a replacement cylinder, scale bar is 100~$\mu$m. {\it B}) Composition along the axis of the cylinder versus distance from first replacement front $I_1$. {\it C}) Cl concentration field from microprobe measurement (intensity proportional to Cl concentration). The positions of the disk-like cavities separating adjacent layers are indicated by the dashed red lines. {\it D}) Composition of the different layers inside a replacement cylinder \emph{versus} time. {\it E}) Composition profile over the whole replacement crystal at different times. The whole profile displays composition jumps from $X_{KBr} \simeq 0.5$ (the composition of the identifiable solid displacement layer nearest to the bulk KCl) to $X_{KBr} = 1$, which correspond to successive replacement layers. $I_1$, $z=0$, is chosen as the origin of the $z$-axis.} 
\label{fig:microprobe} 
\end{figure}

\emph{In situ} time lapse images and \emph{ex situ} optical images were used to determine a number of spatiotemporal relationships (see Figs.~\ref{fig:interface},~\ref{fig:microprobe}).

The growth of the thickness of the replacement material $L(t)$, the distance from the bulk solution to front $I_1$ (the leading boundary of the replacement zone) can be represented reasonably well by $L(t)=\sqrt{D\cdot t}$, with $D~\simeq$~1.7$\cdot 10^{-10}$m$^2$/s. This effective diffusion coefficient is an order of magnitude smaller than the diffusion coefficients of simple univalent electrolytes such as KBr and KCl in water at room temperature, and it is consistent with the idea that the replacement of KBr by KCl is limited by the diffusion of solvated K$^+$, Br$^-$ and Cl$^-$ in the tortuous pore space inside the K(Br,Cl).
Although the central pore associated with each replacement cylinder is almost straight, the cross-section of the channel is much smaller than the area of the KBr crystal exposed to the liquid phase along the leading front, I$_1$. Consequently ion diffusion between the bulk KCl solution and the KBr is reduced relative to diffusion in a bulk solution. The ion diffusing between the bulk KCl solution and the KBr are diverted into the disk-like replacement cavities and the porous replaced crystal between trailing front I$_2$ and the bulk KCl solution. 

Longitudinal composition profiles of the solid phase from the contact with the  bulk solution ($z=-L(t)$) to the front $I_1$ (the leading boundary of the replacement zone, $z=0$),  obtained from electron microprobe analysis, are shown in Fig.~\ref{fig:microprobe}. Measurements performed at different times can be scaled quite well onto a common curve by dividing the positions by $L(t)$. The solid composition varies from $X_{KBr} = 1$ at the reaction front $I_1$ to $X_{KBr} \approx 0$ at the contact with the bulk solution.
A jump in composition from $X_{KBr} = 1$ to 0.72 occurs across the first solution filled disc-like replacement cavity. A detailed composition profile over one replacement cylinder showed that each replacement layer within a cylinder has a constant composition. These compositions are denoted by $X_{KBr}^1$, $X_{KBr}^2$, ... (from bulk KBr to bulk solution), and $X_{KBr}^0=1$. The composition jumps from one replacement layer to the next across successive replacement cavities, and the largest jump is across the cavity adjacent to the original crystal. Each replacement layer moves by dissolution on one side and growth on the other, but their compositions do not change (Fig.~\ref{fig:microprobe}{\it D}). This observation is important because of the insight that it provides into the replacement mechanism. Dissolution and precipitation are coupled by transport of K$^+$, Br$^-$ and Cl$^-$ across the thin solution filled replacement cavities, but they do not occur in the same place. Due to the fluid composition, dissolution occurs on the rightmost part (Figs.~\ref{fig:microprobe}{\it AC} ,~\ref{fig:optical}{\it CD}) of the replacement cavity, and crystal growth occurs on the leftmost boundary. 
This can be seen on time-lapse images (Video S2): Dissolution on one side of the replacement cavity is coupled with precipitation on the other side of the cavity via diffusion of ions across the cavity, and the coupling of precipitation and dissolution, across the fluid filled cavities,  explains why the replacement cylinders appear to translate. Each compositional layer moves by precipitation on one side and dissolution on the opposite side. To maintain a more-or-less constant composition in the solid replacement layers as they advance towards the KBr crystal Br$^-$ ions must diffuse through the central pores towards the bulk KCl, and chloride ions must diffuse through the central pores towards the bulk KBr. 

Seen from view 1, each cavity appears to have a roughly circular shape with a radius of $r_{max} = 5-70 \mu$m, which is evaluated from  $r_{max}=\sqrt{A/\pi}$, where $A$ is the area of the cavity as seen in view 1. The depths, $h(r)$, of the first cavities (those closest to the KBr) can be evaluated using the interference fringes observed on Fig.~\ref{fig:interface}{\it A}. The depths are related to the distance $r$ from the centers of the cavities by $h(r)= a (r_{max}-r)$, where $a = 0.0685 \pm 0.0066$ is the cavity aspect ratio (Fig.~\ref{fig:interface}).

By following one cavity for 90 min, we observed that the central channel position is essentially fixed with time (Fig.~\ref{fig:interface}). This illustrates the surprising stability of the shapes generated by the coupled growth and dissolution process.

\subsection{Transient stage}

The cylinders, made of solid layers, each with a hole in its center, separated by thin fluid filled cavities, are formed during an initial transient stage. The major changes (the complex dynamics leading to the formation of the channels and the solid layers with different compositions) occur during the first few minutes, and this is followed by a slow relaxation to the steady state cylindrical shapes, which takes about 30 min. Here we describe the very early part of the transient, which is the key to understanding the stable pattern observed in the steady state, based on time lapse video images in view 1 with reflected light (Video S3).

\begin{figure}[htb!]
\setlength{\unitlength}{1cm} 
\centering
\begin{picture}(13,8.9)(0,0.0)
\put(0,0){\resizebox{!}{8.9cm}{\includegraphics{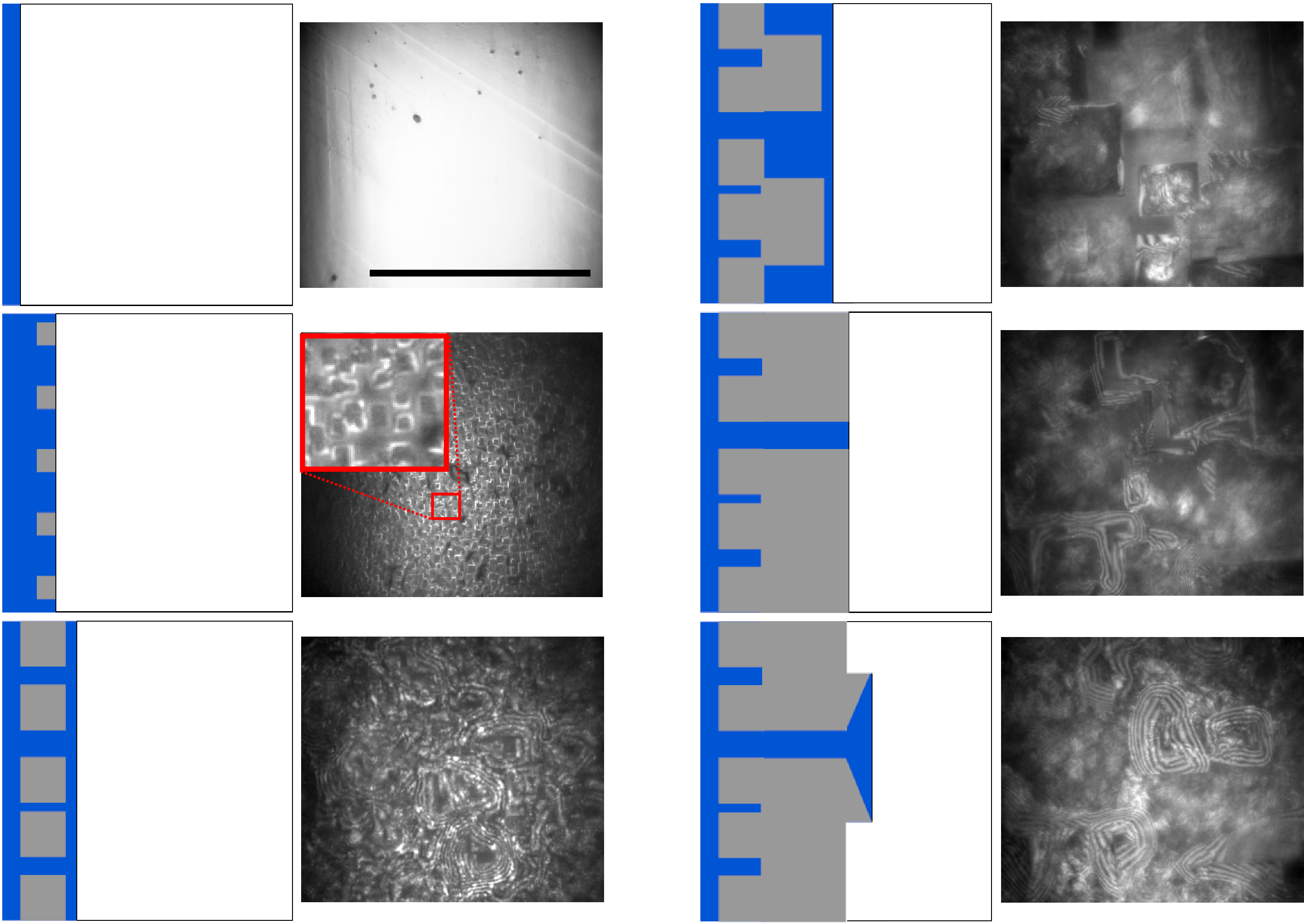}}}
\put(5.4,8.4){\colorbox{white}{A}}
\put(5.4,5.4){\colorbox{white}{B}}
\put(5.4,2.45){\colorbox{white}{C}}
\put(12.15,8.4){\colorbox{white}{D}}
\put(12.15,5.4){\colorbox{white}{E}}
\put(12.15,2.45){\colorbox{white}{F}}
\put(0.1,8.4){\colorbox{white}{\large{S}}}
\put(2.2,8.4){\colorbox{white}{\large{O}}}
\put(0.55,5.4){\colorbox{white}{\large{R}}}
\put(7.5,4.3){\colorbox{white}{\large{$\mathbb{C}$}}}
\put(8.4,1.1){\colorbox{white}{\large{$\mathbb{D}$}}}
\end{picture}
\caption{Transient, early stage, replacement processes. 
Right: Snapshots of the interface (view 1) taken at different times: before immersion ({\it A}), 0.2 ({\it B}), 5 ({\it C}), 60 ({\it D}), 75 ({\it E}) and 115 s ({\it F}). Scale bar is 100~$\mu$m. A close-up of ({\it B}) is displayed. 
Left: 2D representations (unscaled) in view 2 of the view 1 pictures. The colors of the solution, the replaced and original crystals are identical with those of Fig.~\ref{fig:optical}{\it A}. Although the crystals are depicted as separate in the 2D sketch ({\it C}) and ({\it D}) the individual crystallites are connected in the third dimension before detaching from the parent crystal. Symbols refer to the caption of Fig.~\ref{fig:optical}.}
\label{fig:nucleation} 
\end{figure}

In view 1, light is reflected from both the KBr-brine interface closest to the viewer and from the adjacent brine-K(Br,Cl) interface further away (on the opposite side of the fluid filled gap). When these two interfaces are closer than 5-7~$\mu$m, the two reflections interfere and the interference fringes are used to interpret the relative angle and flatness (or waviness) of the two crystal-brine interfaces. Dark regions, the absence of reflection, are due to pores or contact between the two crystal surfaces on opposite sides of the fluid filled gap.

Immediately after the brine is added to the KBr crystal small (7-10~$\mu$m), roughly cubic K(Br,Cl) crystallites grow on its surface. They are all roughly aligned and they cover almost half the KBr surface (Fig.~\ref{fig:nucleation}{\it B}). The shape, alignment and lack of reflection from these crystallites suggest that they grow epitaxially on the KBr surface.

The dissolution front evolves from the space between the crystallites and undercuts the contacts between the original KBr crystal and the K(Br,Cl) crystallites. Meanwhile, the crystallites grow laterally, connect and form larger crystals. The precipitation front now becomes a wavy, continuous surface of interconnected K(Br,Cl) crystals that are detached from the original KBr crystal (Fig.~\ref{fig:nucleation}{\it C}). 

The dissolution front remains wavy while the original KBr crystal and new K(Br,Cl) crystals are out of contact, suggesting that dissolution is transport controlled. However, the growing K(Br,Cl) crystals maintain high index surfaces during growth, and the distance between the two crystal surfaces does not exceed 10~$\mu$m.

As the pores in the growing K(Br,Cl) crystal become narrower and deeper, dissolution and growth slow down and high index surfaces become increasingly dominant. 

A few crystals start growing more quickly than the rest, and they keep up with the dissolving KBr surface while the other crystals lag behind (Fig.~\ref{fig:nucleation}{\it D}).

The more rapidly growing crystals have high index surfaces with misfits of up to 3-4 degrees.  The distances between the centers of these rapidly growing crystals vary from 10 to 100~$\mu$m.  The rapidly growing crystals grow both laterally and towards the original, receding, KBr crystal and end up contacting both the original KBr crystal and each other (Fig.~\ref{fig:nucleation}{\it E}). Open pores are maintained at junctions between the fast growing crystals and persistent thin fluid cavities develop from these pores (Fig.~\ref{fig:nucleation}{\it F}).

Approximately 6.5 min after the beginning of the reaction, replacement cylinders, each with several layers, are well formed and the layer compositions have already reached their asymptotic values (Fig.~\ref{fig:microprobe}{\it DE}).

After this stage, there is almost no additional change in the shape of the channel, but the shape of the cavity still evolves during the next 10-30 min. The discs (immediately behind front $I_1$) then covers approximately 60\% of the plane of the original KBr surface (Fig.~\ref{fig:optical}{\it B}).

\section{Discussion}

The results presented above provide important insights into the mechanisms of solvent-mediated phase transformation involving solid solutions. The formation and maintenance of connected porosity provides transport pathways from the bulk solution to the original crystal. This pore-structure prevents drastic reduction of the replacement rate that would otherwise occur due to formation of a ``protective'' layer of product phase, and it maintains a reactive interface between the precursor crystal and the solution.
The dynamics of the replacement front quickly evolves towards a quasi-steady-state after a transient period, which is the key to the entire replacement mechanism and rate.

During the initial stages of the replacement process, the product phase forms 7-10~$\mu$m crystallites by heterogeneous/epitaxial nucleation and growth on the reactant (KBr) phase. Although a continuous epitaxial film of the product never forms on the reacting surface, this determines the crystal orientation of the product phase throughout the entire growth period by subsequent epitaxial precipitation on the crystallites, which grow to become the precipitated phase. Thus, although most of the newly formed crystal has never been in contact with the parent phase (this is true in the transient stage, but also in the steady state since fluid-filled cavities lie between parent and new formed crystals), its lattice orientation is dictated by that of the parent phase.
In this way, the strain energy associated with epitaxial growth across an interface of significant lattice mismatch remains small.

The formation of a porous network, which secures effective transport to the reactive surface, is sustained in this system, due to the overall reduction in solid volume associated with the replacement reaction (at least $\sim$13\% (Putnis \& Mezger 2004; Pollok 2004)).
The early channel formation occurs because some crystallites grow faster than others in the brine filled space between the original crystal and the replacing crystal (Fig.~\ref{fig:nucleation}{\it D}). This closes the space between replacing crystals except where these crystals merge (Fig.~\ref{fig:nucleation}{\it E}). This competition between the replacing crystals would have to be addressed in detail to understand the transition from the early stage growth of faceted crystals to the more complex growth of porous morphologies in natural systems and elucidate which minerals and conditions will result in a replacement cylinder structure similar to that displayed by KBr-KCl at room temperature, and which will evolve in other ways.
After the first disc-like cavity has been formed (Fig.~\ref{fig:nucleation}{\it F}) ``nucleation'' and development of the second and subsequent cavities could not be directly observed. We speculate that the formation of the second and subsequent cavities is similar to the formation of the first cavity, by nucleation of crystallites onto the preceding layer, but confirmation of this hypothesis would require a quite different experimental approach.

In the steady state, there seems to be two main reasons why the pores remain open in roughly the same position: There is a solid volume deficit in the replacement of KBr by K(Br,Cl) and the misfit (up to 3-4 degrees) between the crystallites that grow to become pore walls hinders complete closing of pores and stabilizes their positions since this maintains flat high index surfaces and straight pores.  Straight pores maximize the rate of solute transport, and the rate at which  the replacement front advances. 
The range of the composition profiles is expected for a transport-limited reactive system with KBr-rich crystals in the leading part of the replacement zone and KCl-rich crystals close to the bulk solution (and far from the KBr). 
Indeed, the solution becomes richer in Br and poorer in Cl when it penetrates the porous material, and assuming that the precipitated solid is locally in near equilibrium with the solution, a solution richer in Br is expected to precipitate a solid that is richer in Br (from the diagram of Fig.~\ref{fig:lippmann}). But detailed profiles show very distinct composition layers, at least inside the replacement cylinders (the fine-porous material formed behind the advancing KBr replacement zone is not resolved with the present setup and the variations in composition in the wakes of the cylinders is too small to discriminate), which are due to the discrete nature of the replacement that occurs between the two sides of a fluid filled cavity. 
The composition of each layer was shown to be stable over time as the front progressed into the crystal, suggesting that epitaxial growth is the main mechanism that occurs on the precipitation side of a fluid filled cavity. This suggests that atomic layers are deposited onto the replaced crystal with exactly the same composition and lattice properties. This prevents a mismatch between the substrate and the deposit that would lead to a strain energy cost. 
This mechanism is maintained by the specific shape of the fluid-filled cavity (with a thickness that is significantly smaller than its radius (Fig.~\ref{fig:interface})), which facilitates the intimate coupling between dissolution on one side and precipitation on the other side of the cavity. 
This mechanism is also very efficient: The composition of the first replacement layer, after the first cavity, is KBr$_{0.72}$Cl$_{0.28}$ (28\% of the bromine in the KBr has been replaced by Cl), almost 50\% of the Br has been replaced in the fifth replacement layer, and more than 50\% replacement occurs in the sixth replacement layer and the porous K(Br,Cl) layer.
The different layers form after only a few minutes, and assuming that epitaxial growth was efficient at the earliest stages, it can be assumed that the layer compositions are determined by the composition of the crystallites that were first able to nucleate onto the original crystal (or subsequent layers) during the very early stages of the reaction.

The scaling of the composition-position data onto a common curve (Fig.~\ref{fig:microprobe}{\it E}) with the replacement thickness $L(t)$, with $L(t) \simeq t^{1/2}$, is typical of diffusion controlled systems. But once the entire original material has been replaced, the replaced material might ripen or equilibrate which would erase or rearrange the internal porosity. In most cases, the final product will be a transparent, non porous crystal, as observed in the  KBr-KCl-H$_2$O system (Putnis {\it et al.} 2005), or another homogeneous material without a trace of the ``transient'' porosity. Some remnants of the channels might occur as fluid inclusion trails, which are often observed in minerals (Crawford \& Hollister 1986).

\section{Conclusion\label{Conclusion}}

\emph{In situ} and \emph{ex situ} studies of replacement processes in the KBr-KCl-H$_2$O system reveal the development of striking pore structures. The fluid film  between the original and replacement crystals was observed for the first time, and this provides the key to understanding the coupled dissolution and reprecipitation processes in solvent-mediated phase transformation. 
Replacement of KBr by a K(Br,Cl) solid solution that becomes more Cl-rich towards the bulk solution, takes place in an array of discrete more-or-less cylindrical compartments near the advancing KBr dissolution front. A complex and highly anisotropic pore-structure evolves as the front advances.  
The rate at which the front advances is controlled by diffusion in an anisotropic channelized porosity network, which is generated by precipitation of replacement reaction product, with a reduction in solid volume.
Initially replacement is driven by dissolution and the epitaxial growth of replacement product crystals with well defined compositions, influenced by fluid composition and lattice mismatch effects. At later stages, the same mechanisms coupled with solid volume reduction and partial restriction of long-range transport allow a surprisingly stable partially ordered pore structure to form in a solid matrix with large composition changes that are strongly correlated, and mechanistically linked, with the ordered pore structure.
The replacement preserves the gross morphology of the crystal and the crystalline orientation, which are characteristics of such transformations (Putnis \& Putnis 2007), despite the fact that dissolution and reprecipitation occurs with an intermediate liquid film. We have explained how the initial stages of the process may imprint the crystal orientation.
Our findings have major implications for understanding transformation processes in many systems in which reduction in solid volume occurs during epitaxial replacement. 
Based on the results reported here, it is not possible to conclude whether or to what degree the distinctive pore structure observed in the KBr-KCl-H$_2$O system is generic. It appears that volume reduction and epitaxial growth are necessary for the characteristic behavior observed in the KBr-KCl-H$_2$O. However, a large body of additional work will be needed to classify porous solid-solid replacement processes and develop a predictive understanding of the evolution of porous morphologies during solvent mediated replacement reactions. 

\section*{acknowledgements}
C. Raufaste thanks Muriel Erambert for her help with the microprobe. 
Andrew Putnis is thanked for insightful discussions over many years. This study was funded by a Center of Excellence grant from the Norwegian Research Council to PGP.



 \label{lastpage}
 \end{document}